\documentclass[twocolumn,showpacs,amsmath,amssymb,aps,floatfix,superscriptaddress]{revtex4}
\usepackage{graphicx}

\usepackage{mciteplus}

\begin{document}

\title{Numerical comparison of a constrained path ensemble and a driven quasisteady state}

\author{Milo\v{s} Kne\v{z}evi\'c}
\affiliation{Cavendish Laboratory, University of Cambridge, Cambridge CB3 0HE, United Kingdom}
\author{R. M. L. Evans\footnote{corresponding author}}
\email{mike.evans@physics.org}
\affiliation{School of Mathematics, University of Leeds, LS2 9JT, United Kingdom}

\date{13 January 2014}

\begin{abstract}
We investigate the correspondence between a non-equilibrium ensemble defined via the distribution of phase-space paths of a Hamiltonian system, and a system driven into a steady-state by non-equilibrium boundary conditions. To discover whether the non-equilibrium path ensemble adequately describes the physics of a driven system, we measure transition rates in a simple one-dimensional model of rotors with Newtonian dynamics and purely conservative interactions. We compare those rates with known properties of the non-equilibrium path ensemble. In doing so, we establish effective protocols for the analysis of transition rates in non-equilibrium quasi-steady states. Transition rates between potential wells and also between phase-space elements are studied, and found to exhibit distinct properties, the more coarse-grained potential wells being effectively further from equilibrium. In all cases the results from the boundary-driven system are close to the path-ensemble predictions, but the question of equivalence of the two remains open.

\end{abstract}



\maketitle

\section{\label{intro}Introduction}

Equilibrium thermal systems obey the principle of detailed balance \cite{vanKampen}, a physical law concerning the rates of a system's microscopic dynamical processes, which can ultimately be traced to the statistics of the heat bath that supplies noise to the system. A class of far-from-equilibrium system, that shares the Hamiltonian of an equilibrium system, can be defined by a subset of the equilibrium ensemble of phase-space trajectories, conditioned by a finite flux \cite{Evans04,Evans05,Jack10,Chetrite13}. That is to say, those members of the equilibrium ensemble of systems, which exhibit a given flux during some time interval, are defined as belonging to the non-equilibrium constrained-flux ensemble. For such a constrained-flux ensemble, the rates of microscopic processes have been shown \cite{Evans04,Evans05,Baule08,Baule10} to respect physical laws equivalent to (but different from) equilibrium detailed balance, that can ultimately be traced to the statistics of a non-equilibrium reservoir that supplies biased noise to the system \cite{Simha08}.

Such an ensemble is appealing in that it shares many features of an equilibrium ensemble, and admits elegant techniques for investigation of it properties, both in the case where the constrained dynamical quantity is antisymmetric under time reversal (a flux) \cite{Evans10CP,Jack10,Chetrite13}, and where it is symmetric (a ``dynamical activity") \cite{Hedges09,Merolle05,Garrahan07,Garrahan09,Chandler10,Jack10b,Speck11,Jack10,Chetrite13}. However, it remains unclear whether such ensembles are realized in practise, i.e.~whether a non-equilibrium ensemble defined via the distribution of its phase-space trajectories in this way, corresponds to a physically realistic system driven away from equilibrium.

Here, we test whether such an ensemble is a good description of a system subjected to torsional shear flow, by investigating relationships between its transition rates $\omega_{ij}$, defined as the probability per unit time that the system occupying microstate $i$ transforms to microstate $j$ within a vanishingly small time interval. For the steady state of an ensemble of trajectories conditioned by a finite mean shear flux, the values of such transition rates are related to the rates $\omega^{\rm eq}_{ij}$ measured in the same fluid (i.e.~with the same Hamiltonian) at equilibrium (in contact with an equilibrium heat bath and not constrained to flow). The relationships \cite{Baule08} are as follows.
\begin{itemize}
\item The product of forward and reverse transition rates between any two microstates is the same in the equilibrium and sheared ensembles, i.e.
\begin{equation}
\label{product}
	\omega_{ij}\,\omega_{ji}=\omega^{\rm eq}_{ij}\,\omega^{\rm eq}_{ji}	
	\qquad\forall\;i,j.
\end{equation}
\item The exit rate (i.e.~the sum of all outward transition rates) from any given microstate differs from its equilibrium value by a shear-rate-dependent constant that is the same for all microstates, i.e.
\begin{equation}
\label{sum}
	\sum_j \left( \omega_{ij}-\omega^{\rm eq}_{ij} \right)=Q\qquad\forall\;i.
\end{equation}
If, in addition to the mean flux, the mean potential energy $\langle U\rangle$ (a time-reversal symmetric quantity) is also constrained, further conditioning the non-equilibrium ensemble of trajectories, then Eq.~(\ref{sum}) no longer holds for all microstates $i$, but continues to hold for all microstates of equal potential energy \cite{Chetrite13}.
\end{itemize}

Testing the applicability of this non-equilibrium counterpart to detailed balance requires the rates of transitions between microstates to be measured, in a suitably driven system that has physically valid equations of motion (rather than one in which non-equilibrium transition rates are specified {\it a priori}). Such a model system --- a line of angular-momentum-conserving rotors with nearest-neighbour interactions --- was simulated in Ref.~\cite{Evans10}, in the Brownian limit, where frictional and stochastic forces between neighbours dominate over momentum degrees of freedom. In the present study, we test the proposed statistical laws in a deterministic Hamiltonian system with non-trivial interactions, and with no friction or stochastic forces (other than the emergent quasi-random motion of the many deterministic degrees of freedom). This model allows us to investigate a phase-space with non-trivial momenta as well as positional coordinates. 

Because the model has continuous degrees of freedom, the ``microstates" that we analyse are, by necessity, not strictly single states, but finite regions of phase space, which we define in two different ways. In the first test (section~\ref{test1}), we assume that momentum degrees of freedom are well thermalized and therefore irrelevant, while positional degrees of freedom make distinct transitions between minima of a potential that play the role of microstates. In the second test (section~\ref{test2}), we divide phase space into a grid of coarse-grained ``microstates" characterised by both positional and momentum coordinates, and study transitions between the cells of that grid. Comparing the two characterisations of the same system --- with and without regard of momenta --- reveals some important issues concerning the statistical mechanics of deterministic systems in non-equilibrium environments.

\section{The model}

\begin{figure}[ht]
\begin{center}
  \includegraphics[width = 8.7cm]{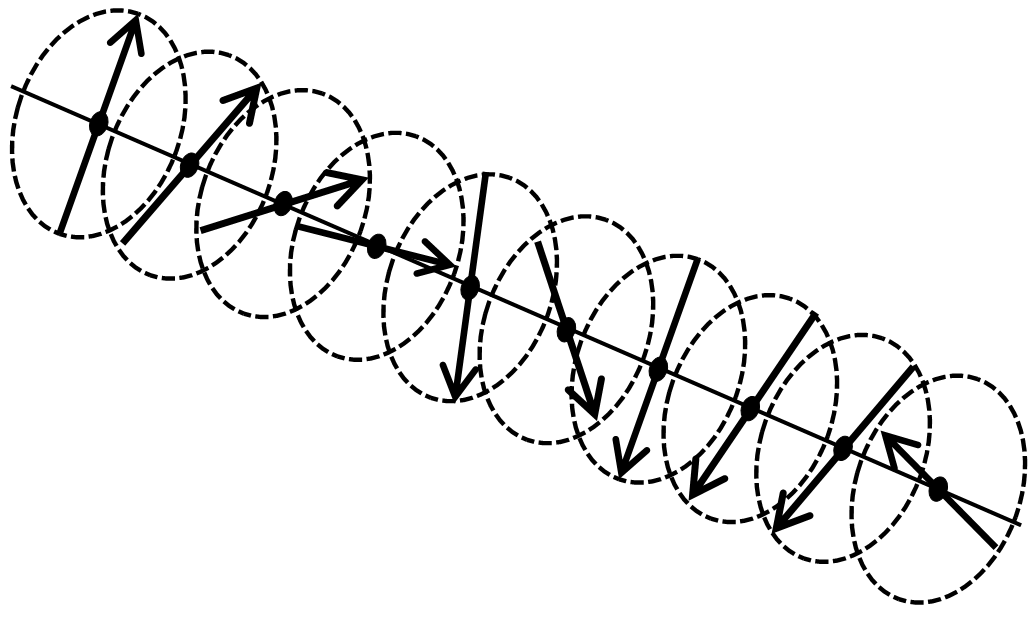}
  \caption{\label{model}\small The model system under investigation: a one-dimensional chain of simple rotors with conservative nearest-neighbour interactions. The chain may be allowed to equilibrate, or can be driven by continuously twisting the boundary rotors. This many-body system is a simple model for a complex fluid in shear flow.}
\end{center}
\end{figure}

The model, in which we measure transition rates for comparison with the predictions for constrained ensembles, is depicted in Fig.~\ref{model}. It consists of a one-dimensional chain of simple rotors. The dynamical variables are the angles $\theta_i$ of the rotors (each labelled by its index $i$) relative to some overall reference direction, and their angular momenta $I\dot{\theta}_i$ where the moment of inertia $I$ will henceforth be set to unity without loss of generality. Nearest neighbours in the chain apply equal and opposite torques to each other, thus exactly conserving angular momentum.

\begin{figure}[ht]
  \includegraphics[width = 8.7cm]{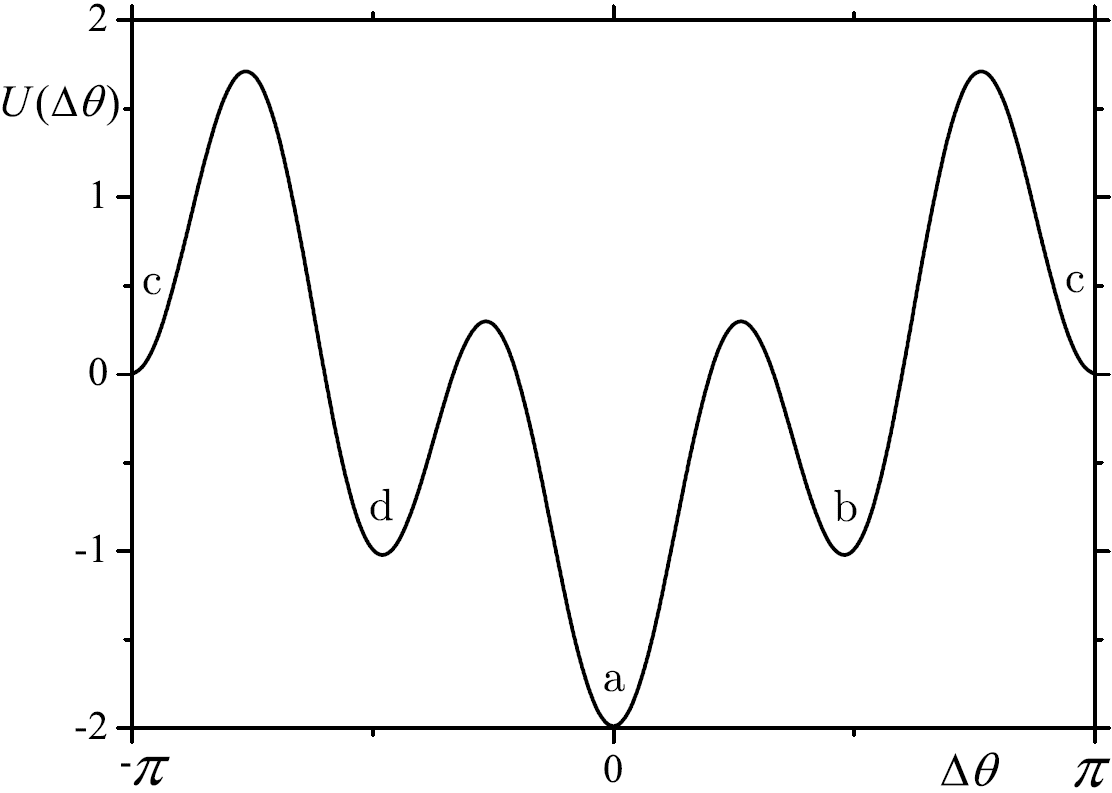}
  \caption{\label{UFig}\small The interaction potential between neighbouring rotors, $U(\Delta\theta)=-\cos(\Delta\theta)-\cos(4\Delta\theta)$, in terms of the angular difference $\Delta\theta$ between the neighbours.\vspace{-4mm}}
\end{figure}

Unlike Ref.~\cite{Evans10}, in which only the zero-mass, over-damped regime with added noise was studied, here we investigate the deterministic model in which the torques are purely conservative, being the negative gradient of the potential $U(\Delta\theta)=-\cos(\Delta\theta)-\cos(4\Delta\theta)$ shown in Fig.~\ref{UFig}. This symmetric function of the angular difference between the rotors, $\Delta\theta_i \equiv \theta_{i+1}-\theta_i$, has four wells, allowing us to measure and compare the transition rates between various states, approximating microstates.

The equations of motion are 
\begin{equation}
\label{EOM}
	\frac{\partial^2 \theta_i}{\partial t^2} = U'(\Delta\theta_i)-U'(\Delta\theta_{i-1}),
\end{equation}
with $U'$ being the derivative of the function $U$.
The boundary conditions are periodic, and the equations of motion are numerically time-stepped using the Velocity Verlet algorithm \cite{VelVerlet}, that approximately conserves energy in the absence of external work. The time step used was $10^{-3}$, as this was found to be sufficiently small to obtain data that were independent of time step. We use a system of $N=300$ rotors, as this is found to be sufficiently large to avoid any system-size dependence of the results.

\begin{figure}[ht]
\begin{center}
  \includegraphics[width = 8.7cm]{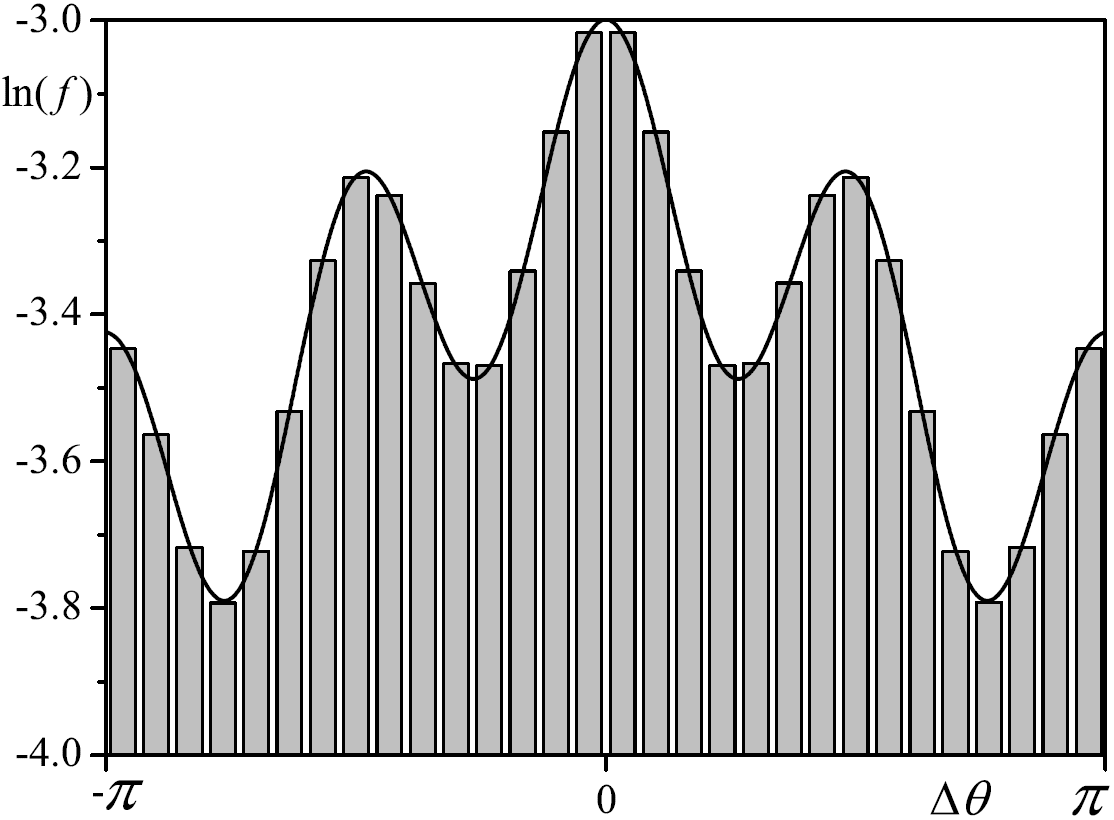}
  \caption{\label{Boltzmann}\small The steady-state distribution of relative angles $\Delta\theta$ between neighbours in the chain without driving. The vertical axis shows $\ln(f)$ where $f$ is the probability per unit angle of finding a given relative angle between neighbours. Also plotted as a continuous curve is $-\beta~U(\Delta\theta)+c$ for the function $U$ given in Fig.~\protect\ref{UFig}, where $\beta$ and $c$ are fitting parameters. Agreement between the curve and the data confirms that the deterministic system is at thermodynamic equilibrium, respecting Boltzmann's law: $f\propto\exp(-\beta U)$.}
\end{center}
\end{figure}

As described thus far, this is an equilibrium model. Once initial transients have died away, the time-averaged distribution of relative angles $\Delta\theta$ between neighbours was measured, and is plotted logarithmically in Fig.~\ref{Boltzmann}, together with the function $-U(\Delta\theta)$ for comparison. Despite having deterministic dynamics, this steady-state system exhibits Boltzmann statistics in the absence of driving. At low temperature (i.e.~evolved from an initial condition with low energy per rotor) the relative angles between neighbours are then mostly confined close to the local potential minima, with only occasional transitions between potential wells. 

The rates of those transitions (without driving) have also been measured, and found to respect detailed balance, thus suggesting that this 1D deterministic system is ergodic, at least in the absence of driving. Note that detailed balance is not imposed \emph{a priori}; it emerges from the dynamics at equilibrium.

To drive the system, in a way reminiscent of shearing a fluid, we modify the boundary condition, which is a simple periodic boundary condition for the equilibrium case. We could simply control the motion of the first and last rotors ($i=1$ and $i=N$), to impose a torsional shear rate, but this might introduce non-trivial edge effects by breaking the translational symmetry. Instead, we impose an angular version of Lees-Edwards boundary conditions \cite{LeesEdwards}, in which rotor $N$ is coupled to rotor $1$ with an angular offset that increases linearly with time, by defining 
\begin{equation}
\label{BC}
	\Delta\theta_{N} = \Delta\theta_0 = \theta_1 - \theta_N + N \dot{\gamma} t
\end{equation}
where the parameter $\dot{\gamma}$ is the shear rate.
So the $N$th rotor interacts with the first rotor via a potential that has a minimum, not at $\theta_1-\theta_N=0$ (i.e.~parallel alignment), but at $\theta_1-\theta_N=-N\dot{\gamma}t$ (an increasing angle). Nevertheless, these boundary rotors could feel a constant interaction force if one of them spins relative to other rotors in the chain. Hence the effect of the boundary condition is non-local, imposing an overall torsional shear flow with average relative velocity between neighbours $\langle\dot{\theta}_{i+1}-\dot{\theta}_i\rangle=\dot{\gamma}$, whilst treating all rotors equally.

\section{Data acquisition from quasi-steady states}

We repeated the deterministic simulations 250 times at each shear rate, with randomly varying initial conditions. The initial angles were set to $\theta_i=0~\forall~i$ in each case (to keep the initial energy low), and the velocities randomly scattered about an affinely sheared state, so that the values of $\Delta\dot{\theta}_i$ have a mean of $\dot{\gamma}$ and a Gaussian distribution whose variance determines the initial energy density. This initial condition is sometimes called a ``water bomb" initialization because, when a child's water bomb (water dropped in a balloon) hits the ground, all of the water is initially in the same location, but with a scatter of velocities that soon cause the water to spread out. We found it convenient to set the variance of the initial Gaussian equal to $\dot{\gamma}/2$, to allow a statistically significant number of the rarest transitions to be observed within a reasonable time, for a range of shear rates. Starting transients were allowed to decay before any data were taken.

Since no dissipation is present, the driven system ``heats up", i.e.~its energy density on average increases with time, since the non-equilibrium boundary condition applies work to the system. This is true also of an adiabatic experimental system. The quasi-steady-state statistics of such a system remain well defined and reproducible so long as it does not heat up too quickly on the time-scale of the measurements.

\begin{figure}[ht]
\begin{center}
  \includegraphics[width = 8.7cm]{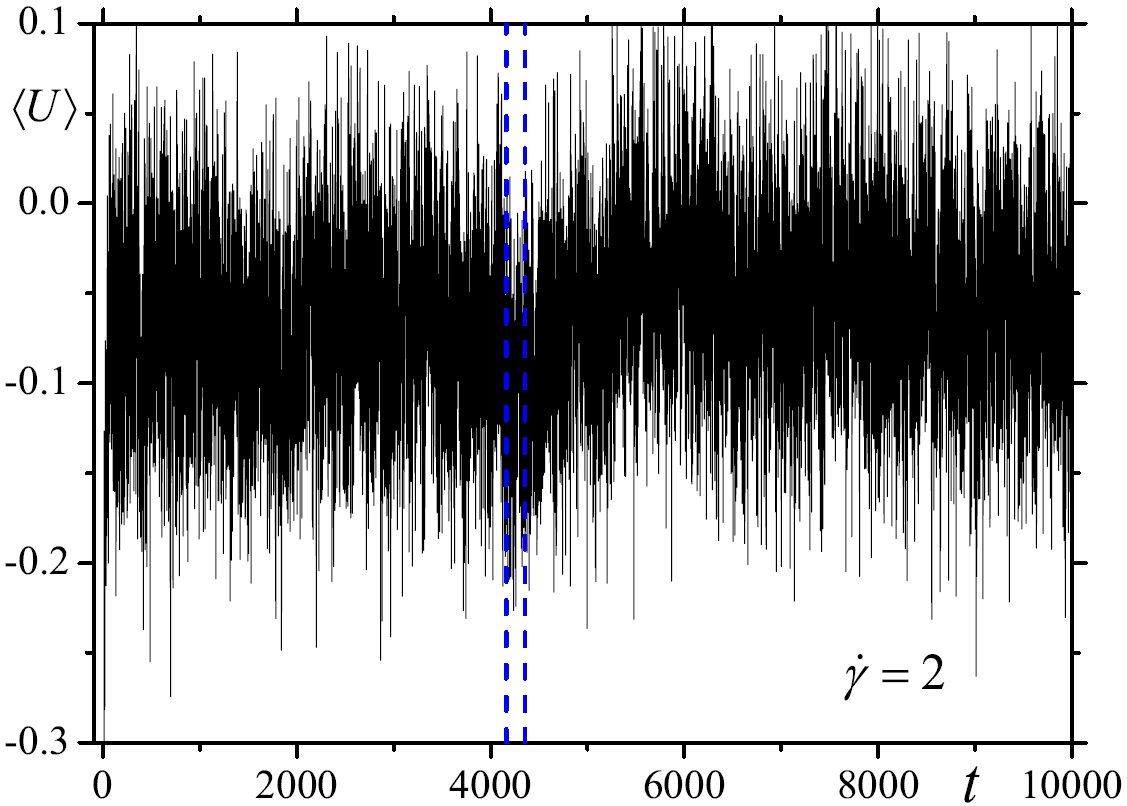}
  \caption{\label{U(t)}\small (color online) Mean potential energy per rotor as a function of time, for one of the many simulations performed at a shear rate $\dot{\gamma}=2$. Vertical dashed lines delineate the interval during which data were obtained with the appropriate mean potential energy per rotor. Note the slow systematic energy increase, only apparent on the longest timescale. Note also the rapid noisy variation on a timescale much shorter than the measurement interval.}
\end{center}
\end{figure}

To measure transition rates at a well defined energy density (corresponding to a constant temperature in the case of an equilibrium simulation), and in a quasi-stationary state, we took data over only short intervals $\Delta\,t$, within each of the much longer simulations, during which there was negligible systematic rise in the energy density compared with its noisy variation (see Fig.~\ref{U(t)}).

\begin{figure}[ht]
\begin{center}
  \includegraphics[width = 8.7cm]{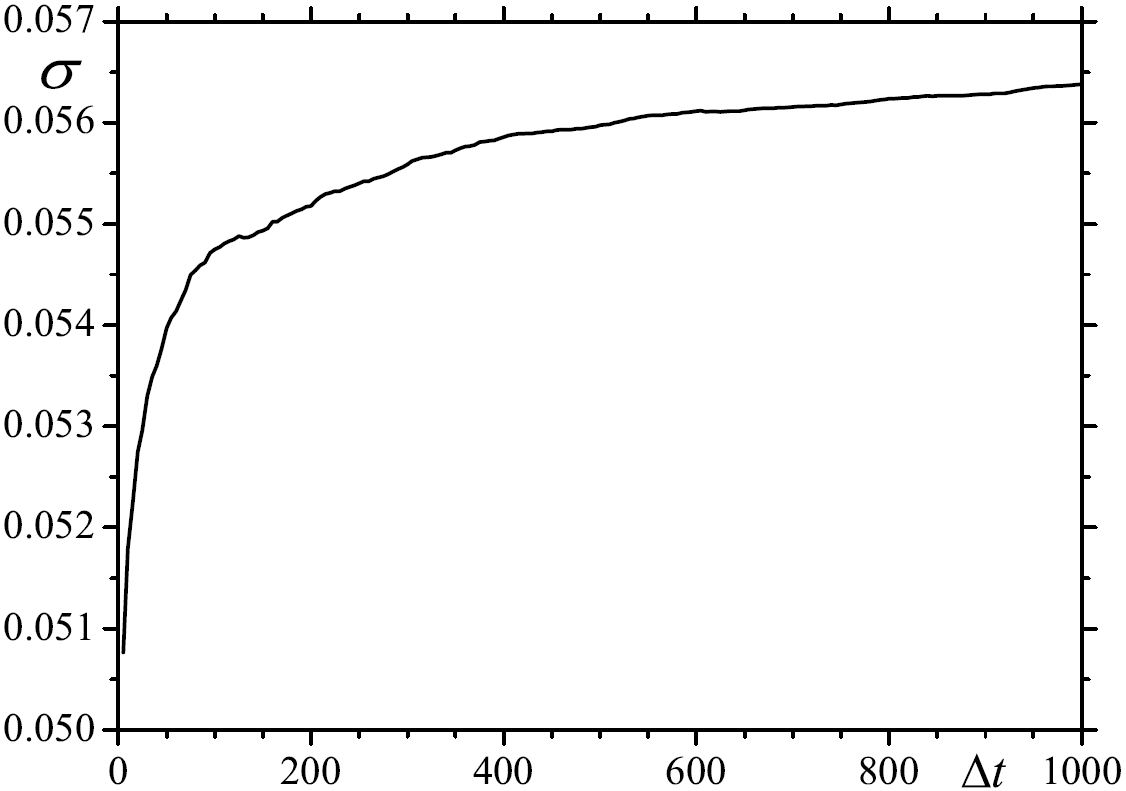}
  \caption{\label{sdeviation}\small Graph of the standard deviation in mean potential energy per rotor (averaged over many simulations) as a function of the measurement interval $\Delta\,t$. The data shown are for simulations at a shear rate $\dot{\gamma}=1$.}
\end{center}
\end{figure}

It is important to choose a value of $\Delta\,t$ longer than the time-scale of the noise but shorter than the timescale on which the internal energy significantly increases in a systematic manner. The data in Fig.~\ref{sdeviation} were used to assess the most suitable choice of $\Delta\,t$. The figure shows a graph of the standard deviation in mean potential energy per rotor (averaged over many simulations) as a function of the measurement interval $\Delta\,t$. The data shown are for simulations at a shear rate $\dot{\gamma}=1$. For other values of $\dot{\gamma}$, the corresponding graphs are qualitatively (and in most cases quantitatively) the same. The graph shows a shoulder, at a timescale required to representatively sample the noisy variations in energy. Further increasing the measurement interval beyond this shoulder value increases the standard deviation only slowly, as it encompasses a systematically increasing range of temperatures (a range of different quasi-steady-states). From the figure, a value of $\Delta\,t$ in the interval $(150,200)$ is judged to be the most suitable quasi-steady timescale. In practice, for each of the investigations presented below, data were collected until the rarest transition, in each case, was observed a given number of times. That number was chosen, for each investigation, to yield values of $\Delta\,t$ in the interval $(150,200)$.

The total duration of each simulation was $10^4$ natural time units, chosen to provide a useful amount of data following starting transients. Within each of the 250 simulated trajectories for each shear rate, several distinct measurements of duration $\Delta\,t$ were taken.

At the end of each of these measurement intervals, counting began again for a new measurement interval. The mean potential energy per rotor was recorded during each interval. Hence, from the many simulations, we obtained a large set of quasi-static measurements with a range of different energy densities. 

To obtain statistically significant data at a well defined energy density, we combined only those measurements for which the mean potential energy per rotor was in the interval $(-0.11,-0.09)$. One of the intervals from which data were used is indicated by the vertical lines in the example of Fig.~\ref{U(t)}. As discussed below Eq.~(\ref{sum}), this extra constraint on the (time-reversal-symmetric) potential energy density only has the effect of confining Eq.~(\ref{sum}) to a set of microstates of equal potential energy density \cite{Chetrite13}.

It was difficult to obtain statistically significant results above a shear rate of $\dot{\gamma}=2$ because the system heated up more quickly at higher shear rate, therefore spending less time in the chosen energy interval.

We choose to characterize the macrostates by their potential energy density because it is well defined, whereas kinetic energy density in the sheared periodic system depends on system size and rest frame. We chose as low an energy as possible, whilst still achieving statistically significant counting, in order to observe non-trivial variations in transition rates, controlled by an interplay between interactions and driving (rather than just a shear-dominated high-temperature regime).

To check that our protocol truly yielded quasi-steady-state results that are independent of initial conditions, we performed detailed comparisons at $\dot{\gamma}=0.4$, $1$, $1.6$ and $2$. For each of these shear rates, we performed a further set of 250 simulations with the initial peculiar velocities drawn from a Gaussian with a smaller variance: a quarter of that in our standard protocol. For these simulations, we re-performed all measurements. In each case, we found that the two sets of results differed by only small amounts, consistent with our quoted uncertainties. We therefore interpret the driven system as reaching an ergodic quasi-steady state, independent of the initial conditions, although the time taken to reach a given value of the internal energy must depend on the initial energy.

The equilibrium state $\dot{\gamma}=0$ is an exception. Because the velocity-Verlet dynamics are approximately conservative, the internal energy of the system remains equal to its initial value, in the absence of driving. Hence the results never become independent of the initial variance of the velocity distribution in this case. For this reason results are given, below, for non-zero values of $\dot{\gamma}$ only.

\section{Results and Discussion}

The potential, shown in Fig.~\ref{UFig}, acts on the variables $\Delta\theta_i$, which are the angular differences between neighbouring rotors. These variables are associated with the gaps, or spaces between rotors. It is the gaps, then, that occupy the potential, and make occasional transitions between its wells, under the influence of the erratic forces from the rest of the system. The rest of the system, then, acts as a non-equilibrium reservoir, that supplies biased noise to a rotor gap, with an average tendency to make $\Delta\theta$ increase with time. If the motion of neighbouring gaps is not strongly correlated, then each gap can be treated as an independent system, weakly coupled to the noisy non-equilibrium reservoir, and driven by it. This situation is directly analogous to an equilibrium system, for which the {\em microstate} (rather than the motion) must be uncorrelated with its surroundings in order for Boltzmann's law to hold and the reservoir to be regarded as weakly coupled.
Subject to the assumption of uncorrelated dynamics, then, we regard each gap as an independent system, with phase-space coordinates $(\Delta\theta,\Delta\dot{\theta})$, for which we can test the statistical laws stated in section~\ref{intro}.

\subsection{\label{test1}Test 1: Transition rates between potential wells}

If transitions between potential wells are rare on the time-scale of temporal correlations in $\Delta\dot{\theta}$ (i.e.~the gap ``forgets" its value of $\Delta\dot{\theta}$ between transitions in $\Delta\theta$), then momenta can be neglected, and potential wells can be regarded as effective microstates, labelled $a$, $b$, $c$ and $d$ in Fig.~\ref{UFig}.

A transition is deemed to have occurred when a gap-angle $\Delta\theta$ is found crossing the {\em bottom} of a potential well, having previously crossed the bottom of another well. This criterion prevents multiple counting due to erratic motion at the threshold of a well. These transition counts are divided by the occupancies of the potential wells, to obtain all eight transition rates between the wells. Note that many transitions were observed, in both the forward and reverse directions, across all four potential barriers. This provides further confidence that the system is in an ergodic state.

\begin{figure}[ht]
\begin{center}
  \includegraphics[width = 8.7cm]{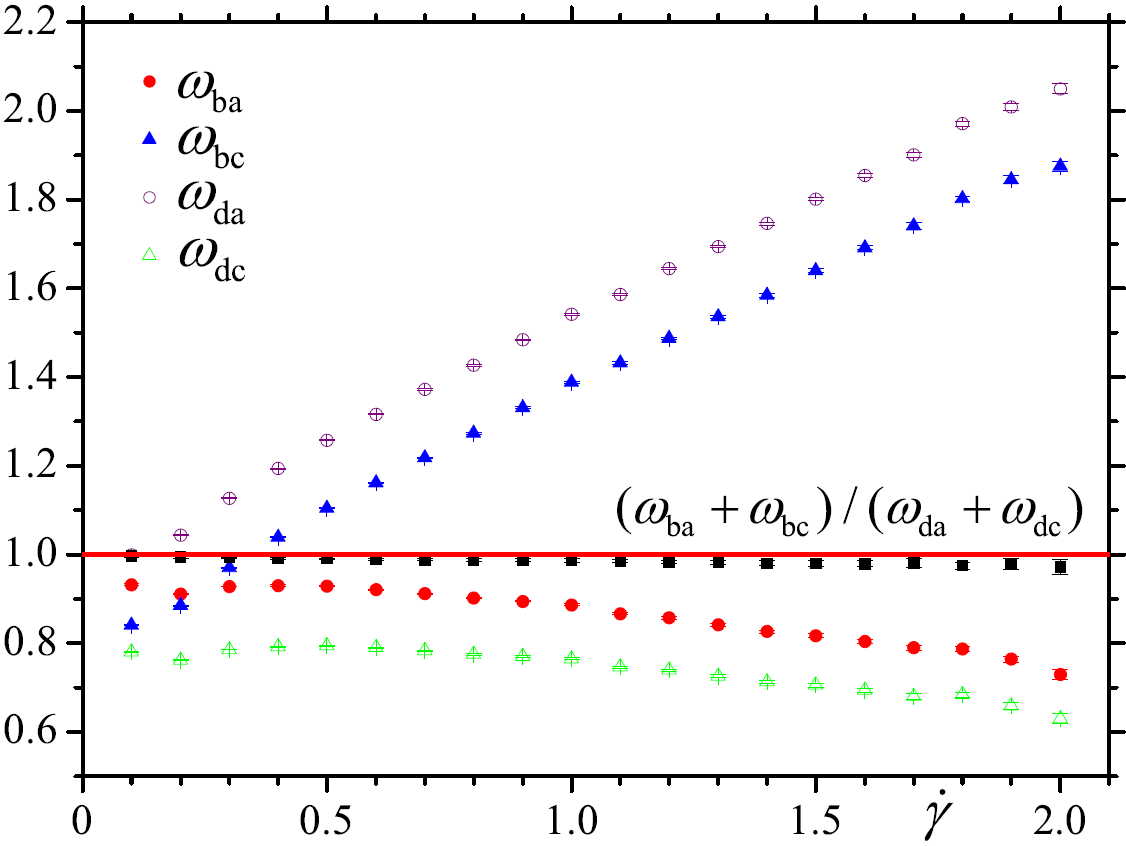}
  \caption{\label{RatesCombo1}(color online) The measured rates $\omega_{\rm ba}$, $\omega_{\rm bc}$, $\omega_{\rm da}$ and $\omega_{\rm dc}$ for a wide range of shear rates $\dot{\gamma}$, both in natural time units. Also plotted is the combination $(\omega_{\rm ba}+\omega_{\rm bc})/(\omega_{\rm da}+\omega_{\rm dc})$ which is the ratio of exit rates from wells b and d, predicted to remain constant at unity in a constrained ensemble, subject to the assumptions that momentum variables can be neglected, and that potential wells represent distinct microstates of a system weakly coupled to the non-equilibrium reservoir embodied by the other rotors. The red line is the theoretical prediction of this ratio (unity), while results of simulations are depicted by black squares. The mean potential energy per rotor is selected in the range $-0.1\pm0.01$.}
\end{center}
\end{figure}

Equation~(\ref{sum}) relates transition rates for the driven system to those at equilibrium. Rather than comparing our driven system with an equilibrium one, for which the temperature would be an unknown fitting parameter, we instead exploit the symmetry of the potential $U(\Delta\theta)$, and notice that the {\em equilibrium} exit rates from wells $b$ and $d$ must be equal, due to the symmetry $\Delta\theta\leftrightarrow-\Delta\theta$ of the equilibrium ensemble. Hence, although that symmetry is broken in the driven ensemble by the shear flux in the positive $\Delta\theta$ direction, substitution into Eq.~(\ref{sum}) nonetheless predicts equal exit rates from wells $b$ and $d$ in the non-equilibrium ensemble. That prediction is tested, in regimes of varying shear rate and internal energy, by the data in Fig.~\ref{RatesCombo1}.

While the four transition rates involved individually behave quite distinctly as shear rate varies, the predicted invariant combination remains almost constant up to large shear rates. So Eq.~(\ref{sum}), which applies to a non-equilibrium ensemble conditioned by flux, appears to be consistent with an ensemble of systems driven by a non-equilibrium shearing reservoir.

\begin{figure}[ht]
\begin{center}
  \includegraphics[width = 8.7cm]{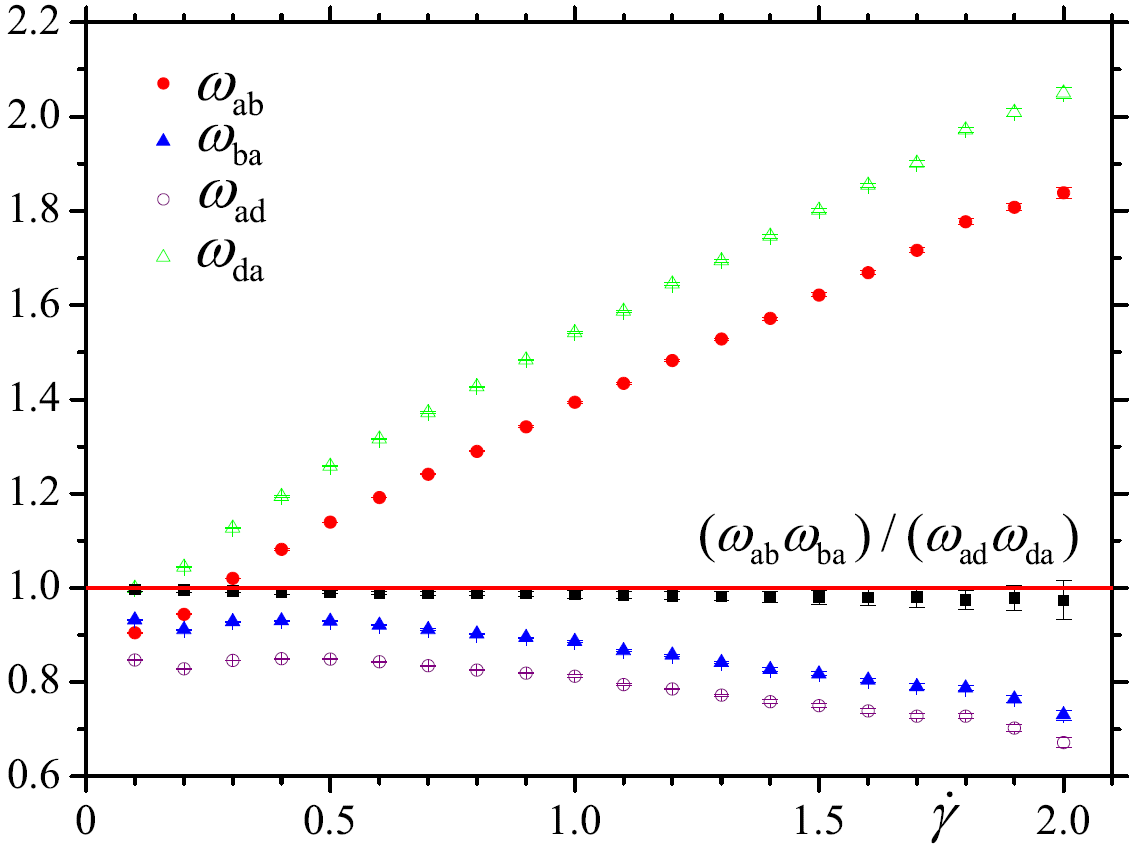}
  \caption{\label{RatesCombo2}(color online) As Fig.~\protect\ref{RatesCombo1}, for the rates $\omega_{\rm ab}$, $\omega_{\rm ba}$, $\omega_{\rm ad}$, $\omega_{\rm da}$ and their combination $(\omega_{\rm ab}\omega_{\rm ba})/(\omega_{\rm ad}\omega_{\rm da})$.}
\end{center}
\end{figure}

\begin{figure}[ht]
\begin{center}
  \includegraphics[width = 8.7cm]{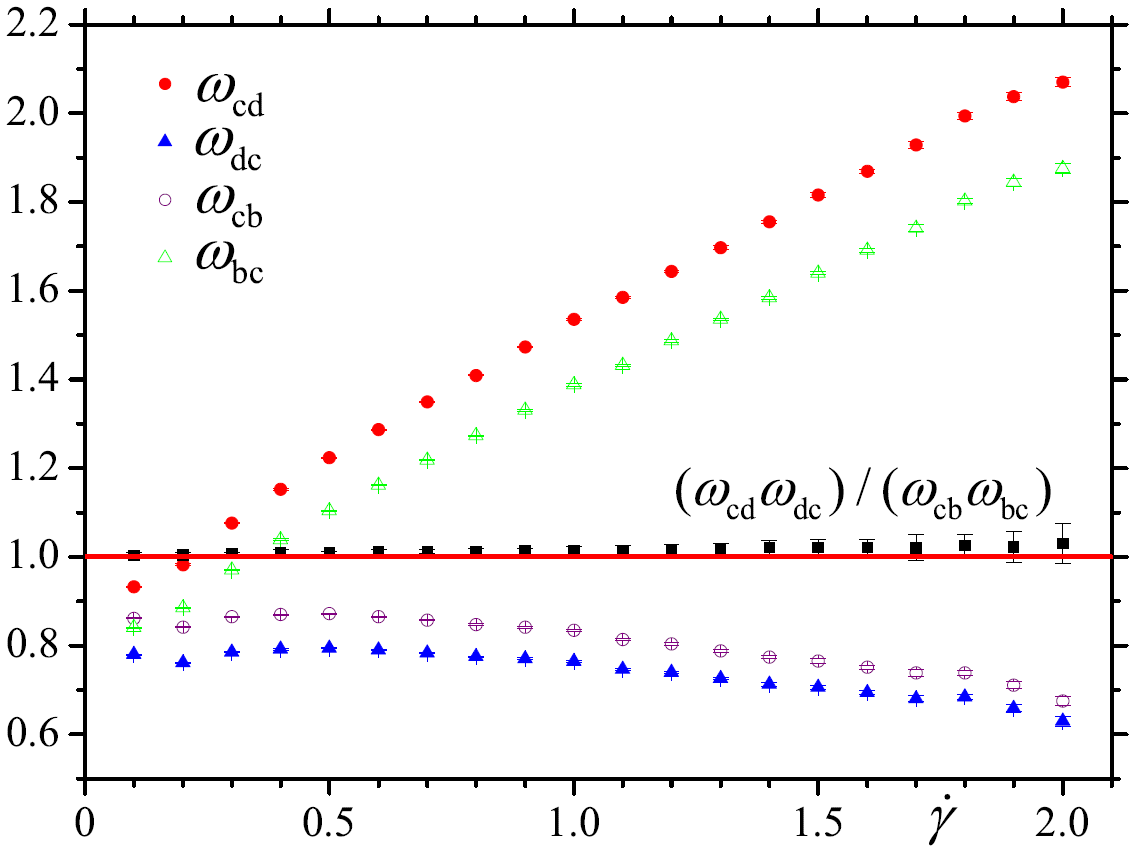}
  \caption{\label{RatesCombo3}(color online) As Fig.~\protect\ref{RatesCombo1}, for the rates $\omega_{\rm cd}$, $\omega_{\rm dc}$, $\omega_{\rm cb}$, $\omega_{\rm bc}$, and their combination $(\omega_{\rm cd}\omega_{\rm dc})/(\omega_{\rm cb}\omega_{\rm bc})$.}
\end{center}
\end{figure}

Again appealing to the symmetry of an equilibrium system with the potential $U(\Delta\theta)$, we test Eq.~(\ref{product}) for the transition from well $a$ to $b$, which must be identical to the transition from $a$ to $d$ at equilibrium, but not when driven. Nevertheless, substitution into Eq.~(\ref{product}) determines that the products of forward and reverse transition rates between these pairs of states remain equal in the constrained non-equilibrium ensemble. That is, 
$\omega_{\rm ab}\omega_{\rm ba}=\omega_{\rm ad}\omega_{\rm da}$ irrespective of the constrained flux. This relation is tested in our driven system by the data in Fig.~\ref{RatesCombo2}, and the relation $\omega_{\rm cd}\omega_{\rm dc}=\omega_{\rm cb}\omega_{\rm bc}$, based on symmetry about well $c$, is tested in Fig.~\ref{RatesCombo3}.
Again, we see that the non-equilibrium invariants remain almost constant while the relevant individual transition rates vary distinctly and by large amounts.


\subsection{\label{test2}Test 2: Phase-space transition rates}

\begin{figure}[ht]
\begin{center}
  \includegraphics[width = 8cm]{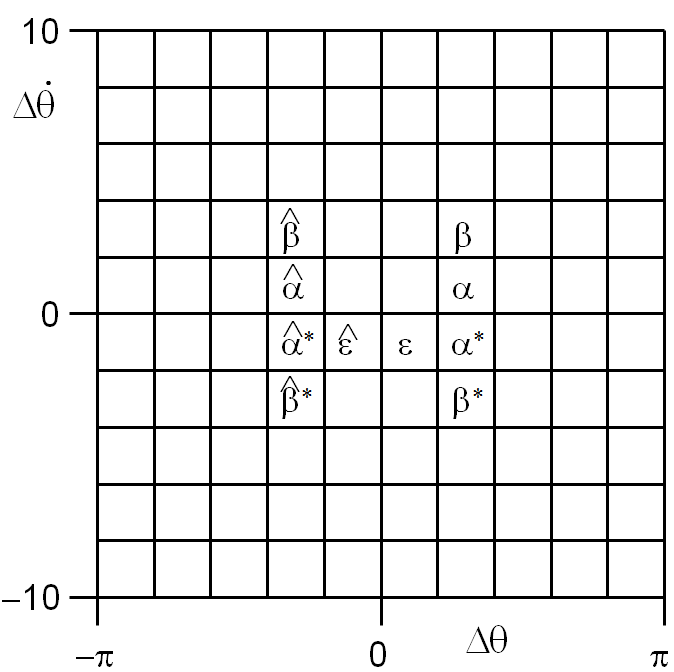}
  \caption{\label{phasespace}Discretization of the phase-space occupied by the gaps between neighbouring rotors, with coordinates $\Delta\theta$ (the relative angle between the pair) and $\Delta\dot{\theta}$ (their relative angular velocity). The states $\alpha$, $\beta$ and $\varepsilon$ are identified for analysis, as well as states reversed in $\Delta\theta$ only, denoted by $\,\widehat{}\,$, and time-reversed states denoted by *.}
\end{center}
\end{figure}

As discussed in section~\ref{intro}, the predicted relations strictly concern the rates of transitions between discrete microstates, and we wish to test them in the continuum limit for a phase space with non-trivial momentum as well as positional degrees of freedom. To that end, we again treat each inter-rotor gap as a system weakly coupled to a non-equilibrium reservoir but, in contrast to section~\ref{test1}, we monitor both its ``positional" coordinate $\Delta\theta$ and its ``momentum" degree of freedom $\Delta\dot{\theta}$. To determine the phase-space occupancies and transition rates, the two-dimensional phase space spanned by these coordinates is discretized as shown in Fig.~\ref{phasespace}. At each time-step during the measuring interval, the values measured in a quasi-steady state are binned into 100 cells (ten columns across a full turn of the periodic coordinate $\Delta\theta$, and ten rows spanning a limited domain of $\Delta\dot{\theta}$ values). A transition is recorded whenever a gap's coordinates cross a line between two such cells.

\begin{figure}
(a) \\ 
  \includegraphics[width = 8.7cm]{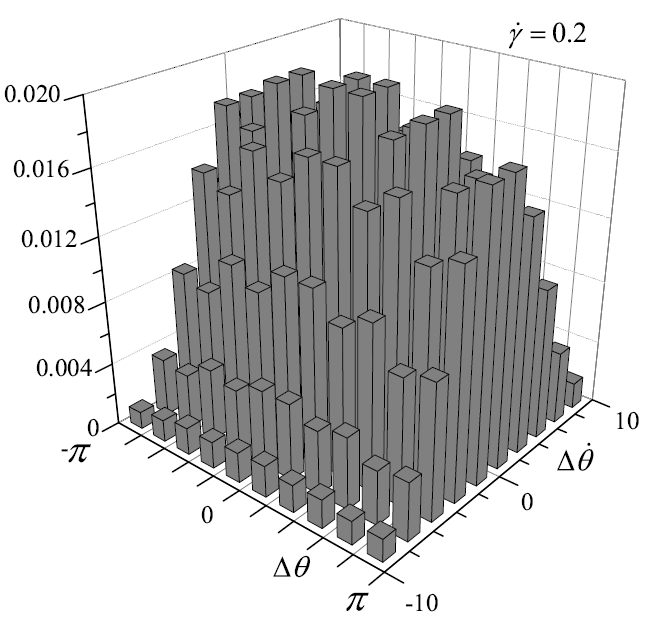}  \\
  \includegraphics[width = 8.7cm]{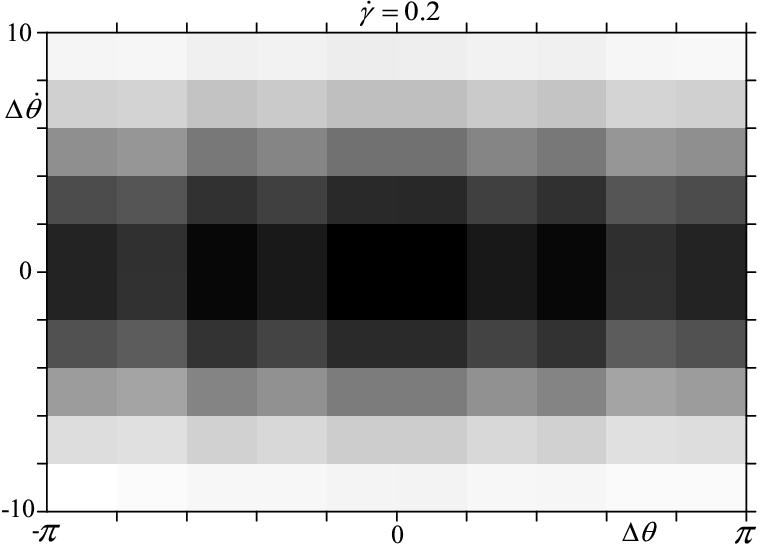}  
  \caption{\label{occupancies} Histograms and density plots of occupancies of the phase-space bins defined in Fig.~\protect\ref{phasespace}, measured in quasi-steady states with mean potential energy density $-0.1\pm-0.01$ and shear rate (a) $\dot{\gamma}=0.2$, (b) $\dot{\gamma}=2$.}
\end{figure}

The measured quasi-steady-state occupancies at two different shear rates, but the same mean potential energy density $-0.1\pm-0.01$, are shown in Fig.~\ref{occupancies}.
At low shear rate (Fig.~\ref{occupancies}a), $\dot{\gamma}=0.2$, the ensemble of rotors is not far from equilibrium, with the angular ($\Delta\theta$) distribution similar to the Boltzmann distribution of Fig.~\ref{Boltzmann} and the velocity ($\Delta\dot{\theta}$) distribution independent of $\Delta\theta$ and approximately Gaussian with a mean value of $0.2$. At shear rate $\dot{\gamma}=2$ (Fig.~\ref{occupancies}b), the velocity distribution has more non-trivial structure, no longer resembling the equilibrium Gaussian form, while the angular distribution is also significantly altered, particularly at the most negative velocity. 

\begin{figure}
(b) \\ 
  \includegraphics[width = 8.7cm]{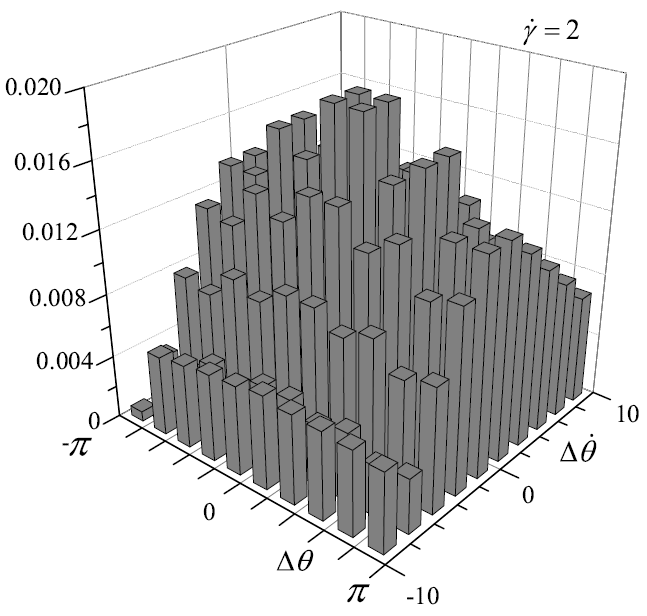}  \\
  \includegraphics[width = 8.7cm]{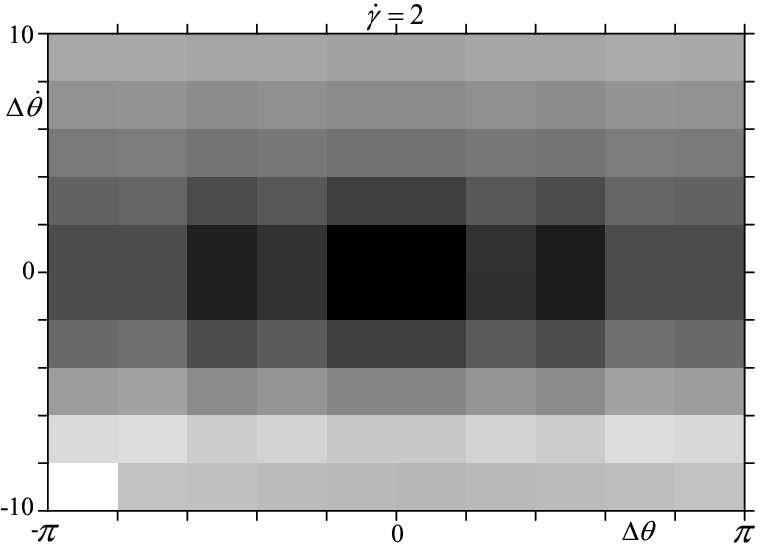}  \\
\end{figure}

\begin{figure}[ht]
\begin{center}
  \includegraphics[width = 8.7cm]{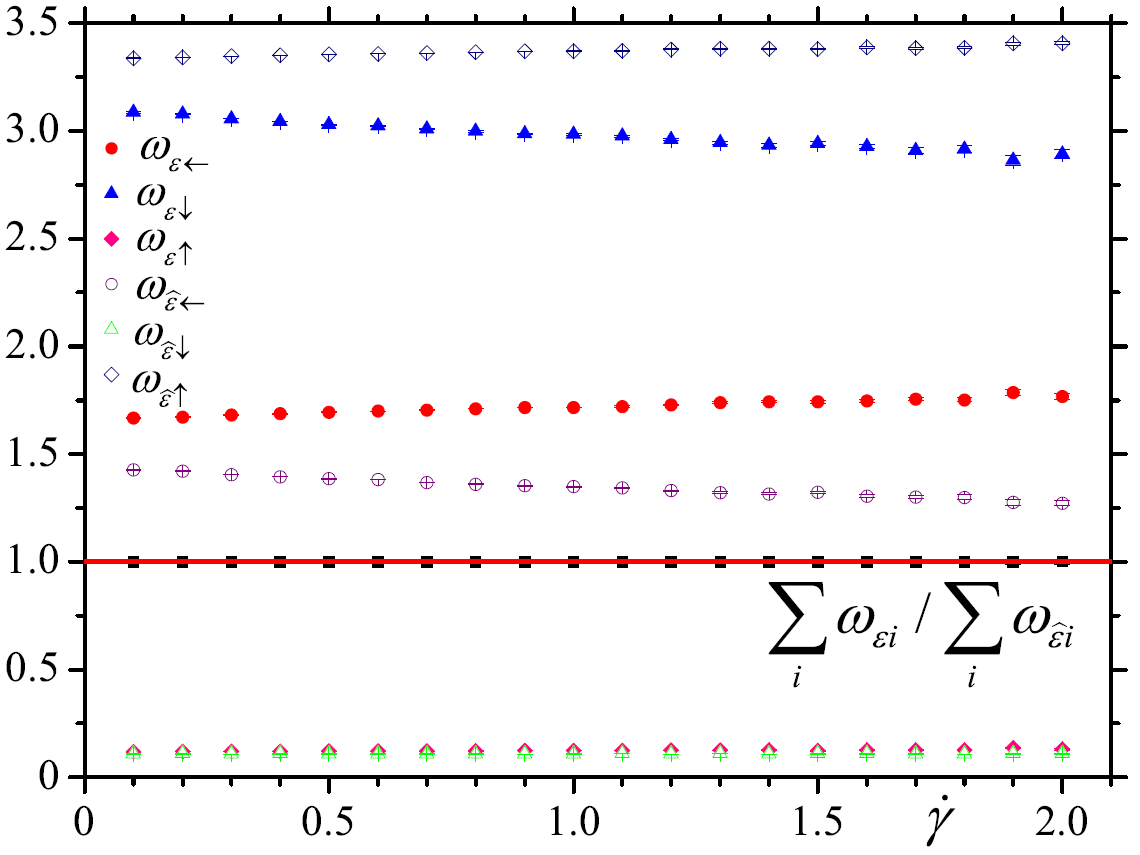}
  \caption{\label{Part2Sum}(color online) Test of the relationship $\sum_i \omega_{\varepsilon\,i} = \sum_i \omega_{\widehat{\varepsilon}\,i}$ between exit rates from states $\varepsilon$ and $\widehat{\varepsilon}$, defined in Fig.~\ref{phasespace}, in quasi-steady states with mean potential energy per rotor $-0.1\pm~0.01$ as for Fig.~\ref{RatesCombo1}. The key gives symbols representing the individual rates $\omega_{\varepsilon\leftarrow}$ etc.~where the arrow indicates the direction (relating to Fig.~\protect\ref{phasespace}) of the transition out of the initial state. Black symbols: $\sum_i \omega_{\varepsilon\,i} \left/ \sum_i \omega_{\widehat{\varepsilon}\,i}\right.$ are in very close agreement with the theoretical value (red line) of unity.}
\end{center}
\end{figure}

The measured transition rates are compared with the relationships that hold for a conditioned ensemble of trajectories, Eqs.~(\ref{product}) and (\ref{sum}). As in section~\ref{test1}, to test the relation between exit rates, Eq.~(\ref{sum}), we appeal to the symmetry of the equivalent equilibrium ensemble. In particular, we concentrate on the cell labelled $\varepsilon$ in Fig.~\ref{phasespace}, and its inverted image $\widehat{\varepsilon}$. Note that these two states are related by the transformation $(\Delta\theta,\Delta\dot{\theta})\leftrightarrow(-\Delta\theta,\Delta\dot{\theta})$, equivalent to parity {\em and} time reversal, ``PT" (since $\Delta\dot{\theta}$ is not reversed). The equilibrium ensemble is invariant under that transformation, so Eq.~(\ref{sum}) implies $\sum_i \omega_{\varepsilon\,i} = \sum_i \omega_{\widehat{\varepsilon}\,i}$, an equation involving eight transition rates in the driven ensemble (transitions into the four neighbors of $\varepsilon$ and the four neighbors of $\widehat{\varepsilon}$). Two of those rates are for forbidden transitions, and therefore vanish both at equilibrium and with driving. They involve escape, in the direction of increasing $\Delta\theta$, from a state with negative velocity, which would require an improbably large stochastic impulse from the reservoir. The remaining six finite rates are plotted in Fig.~\ref{Part2Sum} for various shear rates, for quasi-steady states selected with mean potential energy density $-0.1\pm 0.01$, as in section \ref{test1}. The ratio of the two exit rates (the combination predicted to be invariant at unity) is also plotted.

\begin{figure}[ht]
\begin{center}
  \includegraphics[width = 8.7cm]{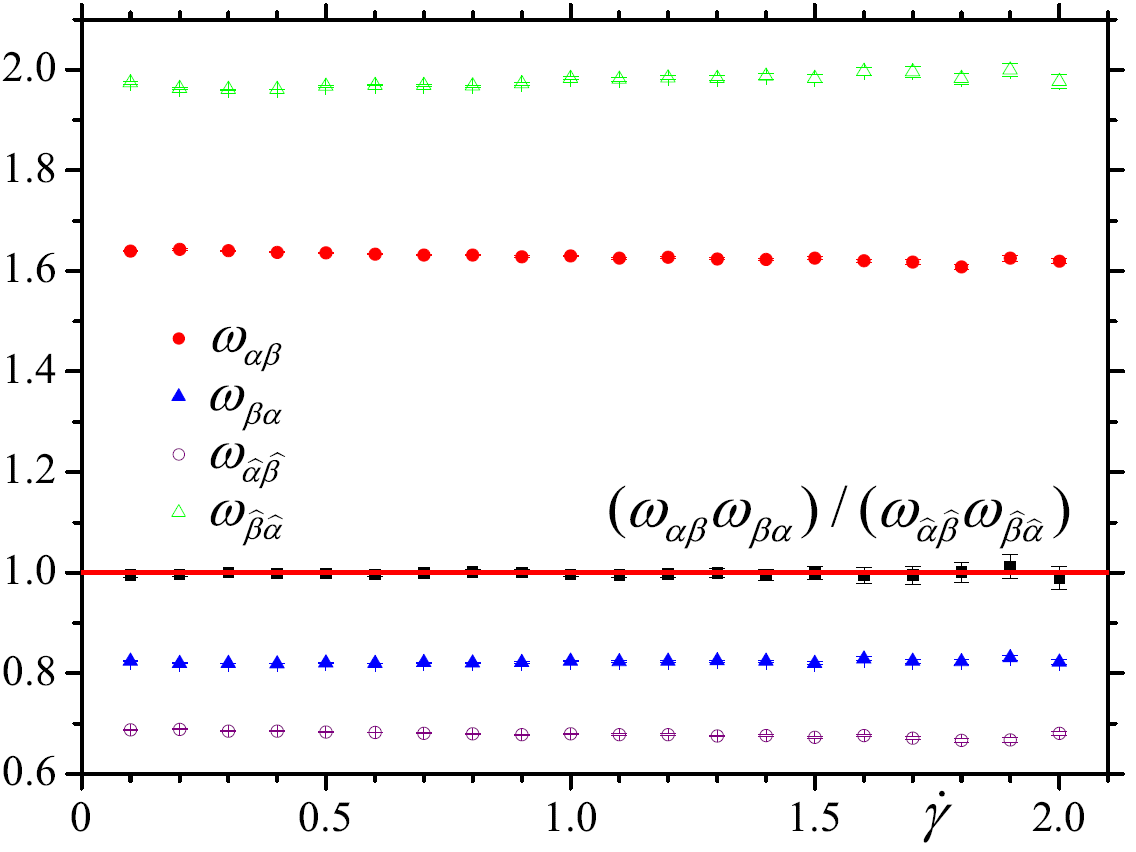}
  \caption{\label{Part2Rel1}(color online) Test of the relationship $\omega_{\alpha\beta}\omega_{\beta\alpha}=\omega_{\widehat{\alpha}\widehat{\beta}}\omega_{\widehat{\beta}\widehat{\alpha}}$ for states $\alpha$ and $\beta$ defined in Fig.~\ref{phasespace}, in quasi-steady states with mean potential energy per rotor $-0.1\pm~0.01$ as for Fig.~\ref{RatesCombo1}. Black symbols: $\omega_{\alpha\beta}\omega_{\beta\alpha}/\omega_{\widehat{\alpha}\widehat{\beta}}\omega_{\widehat{\beta}\widehat{\alpha}}$ are in very close agreement with the theoretical value (red line) of unity.}
\end{center}
\end{figure}

\begin{figure}[ht]
\begin{center}
  \includegraphics[width = 8.7cm]{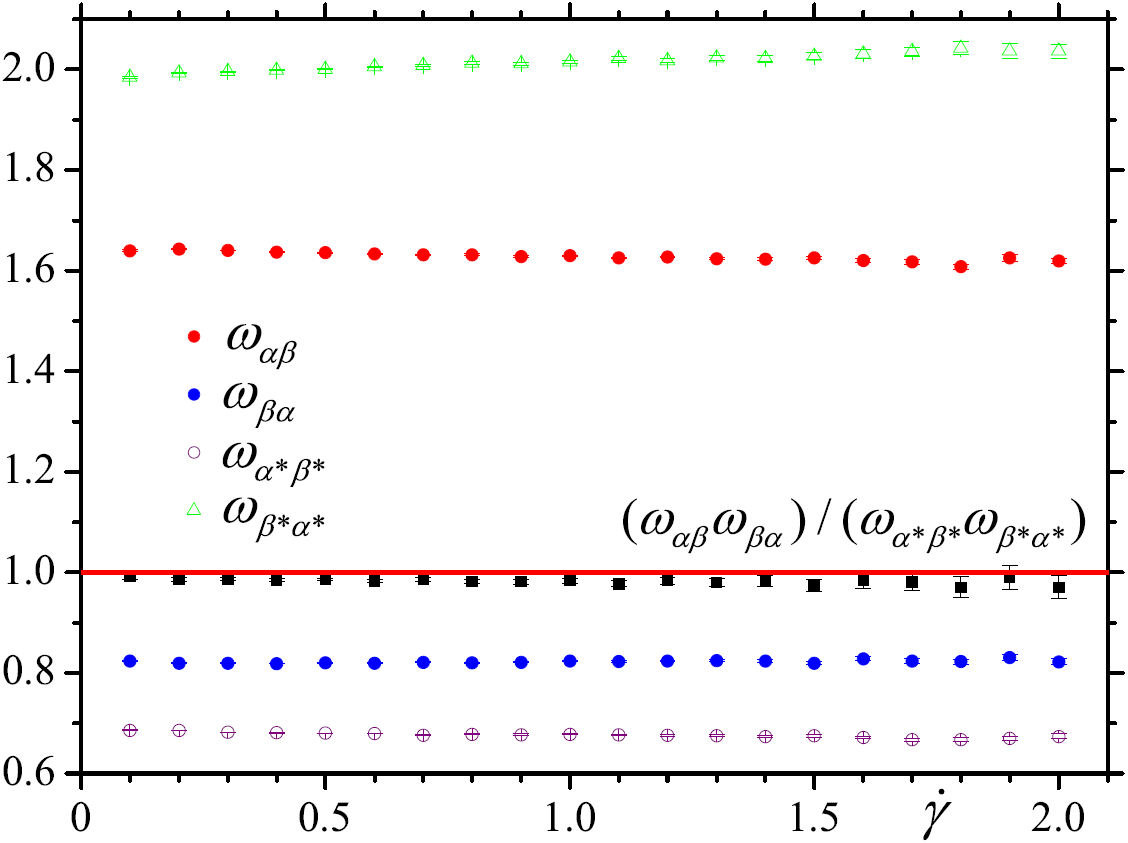}
  \caption{\label{Part2Rel2}(color online) As Fig.~\protect\ref{Part2Rel1}, for the relationship $\omega_{\alpha\beta}\omega_{\beta\alpha}=\omega_{\alpha^* \beta^*}\omega_{\beta^* \alpha^*}$.}
\end{center}
\end{figure}

\begin{figure}
\begin{center}
  \includegraphics[width = 8.7cm]{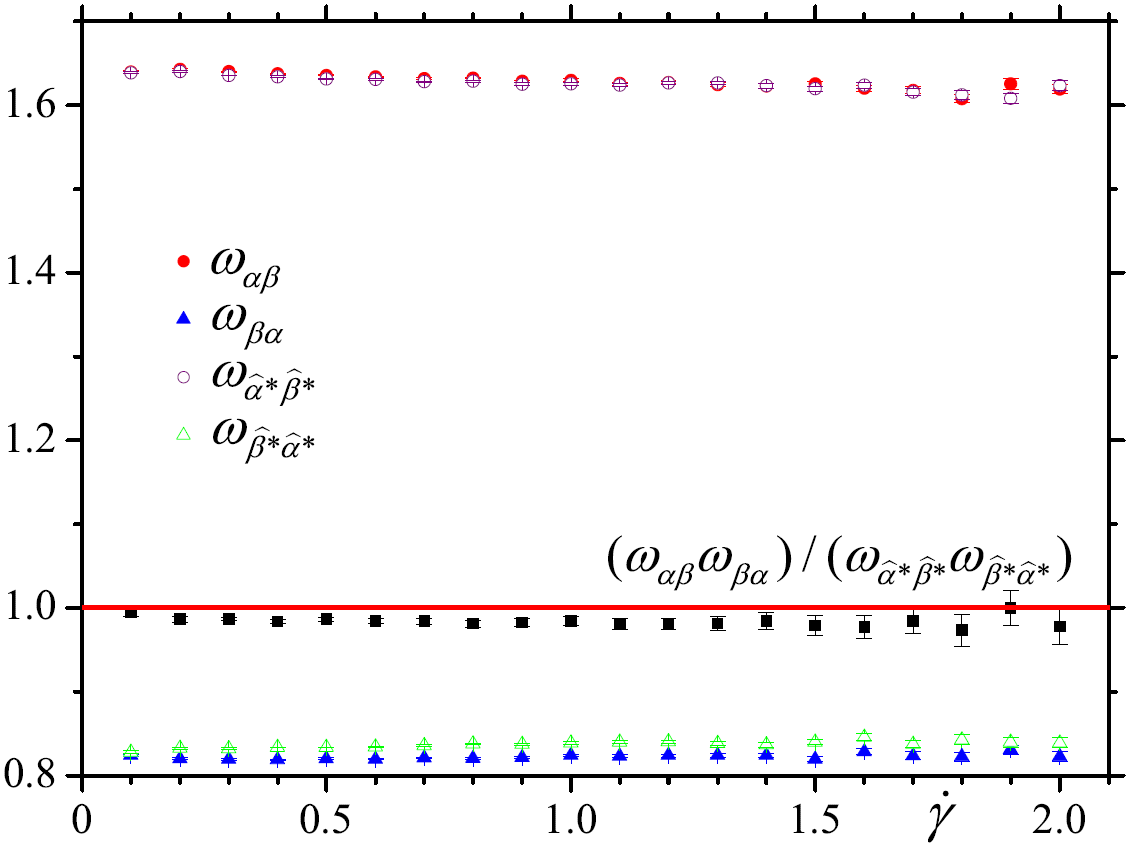}
  \caption{\label{Part2Rel3}(color online) As Fig.~\protect\ref{Part2Rel1}, for the relationship $\omega_{\alpha\beta}\omega_{\beta\alpha}=\omega_{\widehat{\alpha}^*\widehat{\beta}^*}\omega_{\widehat{\beta}^*\widehat{\alpha}^*}$.}
\end{center}
\end{figure}

Similarly, in Figs.~\ref{Part2Rel1}, \ref{Part2Rel2} and  \ref{Part2Rel3}, the product relations in Eq.~(\ref{product}) are compared with data for transitions between ``microstates" $\alpha$ and $\beta$ (see Fig.~\ref{phasespace}) and their symmetry-related microstates, which have equal statistical properties at equilibrium. The symmetries in question are PT, discussed above and denoted by a hat ($\:\widehat{}\;$) and T, simple time reversal, $(\Delta\theta,\Delta\dot{\theta})\leftrightarrow(\Delta\theta,-\Delta\dot{\theta})$ denoted by an asterisk (*). Some small deviations from unity are observed in some cases.


Notice that the transition rates vary much less, across the range of shear rates, than the transition rates measured in section~\ref{test1}. Nevertheless, their small changes are sufficient to result in significantly altered occupancies (Figs.~\ref{occupancies}a \& b). This is because phase space is now being examined in finer detail (Fig.~\ref{phasespace}). The occupancies are the result of many microscopic transition rates, so that tiny changes in each of those many rates, leading to tiny changes in the relative occupancies of neighbouring cells, can have a large effect on the large-scale shape of the distribution.

One might ask whether the ensemble under investigation is near or far from equilibrium. The answer depends on the level of detail at which its properties are interrogated. In terms of transitions between the potential wells, the ensemble at $\dot{\gamma}=2$ is far from equilibrium, as witnessed by the rates in Figs.~\ref{RatesCombo1}, \ref{RatesCombo2} and \ref{RatesCombo3}, which are grossly altered from their equilibrium values (which can be inferred from the graphs by extrapolating to $\dot{\gamma}=0$). The relationships in Eqs.~\ref{product} and \ref{sum} are tested far from equilibrium in that case. However, at a more microscopic level the transitions between ``microstates" in Fig.~\ref{phasespace} have rates that are barely perturbed from their equilibrium values by an imposed shear rate $\dot{\gamma}=2$, as shown in Figs.~\ref{Part2Sum}---\ref{Part2Rel3}, so that the relationships (Eqs.~\ref{product} and \ref{sum}), between the equilibrium and driven rates trivially hold (approximately) in this case.

This scale-dependence of the distance from equilibrium is a very general phenomenon. For instance, when polymeric fluids flow, the constituent polymer-chains become stretch into highly non-equilibrium conformations on the large scale. But more detailed measurements of their small-scale conformations remain equilibrium-like up to much higher flow rates. The system exists both near to and far from equilibrium, depending on the scale of the properties being measured \cite{Doi+Edwards}.

\section{Conclusion}

It remains the case that all of the quantities tested, that are invariant in a constrained ensemble, are found to lie very close to unity in all of the cases that we have measured in the quasi-steady states of our sheared deterministic system. This is a non-trivial observation, but not sufficient in itself to conclude that the boundary-driven set of coupled systems are in every way consistent with the hypothetical conditioned ensemble of trajectories. It is unclear whether the very small deviations from unity observed in Figs.~\ref{Part2Rel2} and \ref{Part2Rel3} arise from a discrepancy between the two types of ensembles, or only from the non-ideality of the microstates approximated by finite bins in Fig.~\ref{phasespace}.

We speculate that this one-dimensional model captures some of the essential physics of real complex fluids in steady shear flow. In such fluids, one of the three dimensions is assigned a special role by the imposition of a velocity gradient, while the other two dimensions contribute only to the complexity of the interactions. Nevertheless, it would be interesting, in future work, to study our driven rotor model on a higher-dimensional lattice, to observe the effects of added dimensions perpendicular to the gradient.

\section{Acknowledgments}

We are grateful to Adrian Baule for helpful discussions. This project was supported by funding from the International Association for the Exchange of Students for Technical Experience (IAESTE).

\end{document}